# Fast and accurate waveform modeling of long-haul multi-channel optical fiber transmission using a hybrid model-data driven scheme

Hang Yang, Zekun Niu, Haochen Zhao, Shilin Xiao, Weisheng Hu and Lilin Yi*

*Abstract*—The modeling of optical wave propagation in optical fiber is a task of fast and accurate solving the nonlinear Schrödinger equation (NLSE), and can enable the optical system design, digital signal processing verification and fast waveform calculation. Traditional waveform modeling of full-time and full-frequency information is the split-step Fourier method (SSFM), which has long been regarded as challenging in long-haul wavelength division multiplexing (WDM) optical fiber communication systems because it is extremely time-consuming. Here we propose a linear-nonlinear feature decoupling distributed (FDD) waveform modeling scheme to model long-haul WDM fiber channel, where the channel linear effects are modelled by the NLSE-derived model-driven methods and the nonlinear effects are modelled by the data-driven deep learning methods. Meanwhile, the proposed scheme only focuses on one-span fiber distance fitting, and then recursively transmits the model to achieve the required transmission distance. The proposed modeling scheme is demonstrated to have high accuracy, high computing speeds, and robust generalization abilities for different optical launch powers, modulation formats, channel numbers and transmission distances. The total running time of FDD waveform modeling scheme for 41-channel 1040-km fiber transmission is only 3 minutes versus more than 2 hours using SSFM for each input condition, which achieves a 98% reduction in computing time. Considering the multi-round optimization by adjusting system parameters, the complexity reduction is significant. The results represent a remarkable improvement in nonlinear fiber modeling and open up novel perspectives for solution of NLSE-like partial differential equations and optical fiber physics problems.

## I. INTRODUCTION

Optical fiber communication forms the main infrastructure of modern communication systems. The low loss and large bandwidth characteristics of optical fibers support long-haul transmission of massive quantities of data [1, 2]. The theoretical basis of optical fiber communication involves study of light transmission characteristics in optical fibers. Fiber channel modeling, which is dominated by use of the nonlinear Schrödinger equation (NLSE) [3], is highly significant for design and simulation of optical fiber communication systems. Accurate optical fiber channel models can aid in understanding of nonlinear dynamics [4], the effects of signal transmission [5,6] and the capacity boundary [7] in optical fibers towards realization of optimal optical fiber communication system structure designs. Additionally, fast and accurate optical fiber channel simulations help researchers to evaluate communication algorithms quickly [8, 9] and undertake performance predictions [10, 11], thus breaking through the physical limitations caused by requirements for expensive instruments and devices and enabling progress in optical fiber communication research.

Traditional fiber channel waveform modeling is mainly based on the split-step Fourier method (SSFM) [3], which is a numerical method for solution of the NLSE. This method requires the total optical fiber transmission distance to be divided into multiple steps using a small segment distance, and the linearities and nonlinearities are calculated separately for each transmission step. Use of a shorter step distance provides higher modeling accuracy but also requires longer computing times. To guarantee high accuracy, the computational complexity must be sacrificed, particularly for large-scale wavelength division multiplexing (WDM) long-haul fiber transmission [12], which is a typical optical fiber communication approach. SSFM takes a few hours to model long-haul WDM fiber channels for a single input condition. For the optimal system design, multi-round optimizations are required by adjusting system parameters, which will take a few days therefore limiting its practical applications. The ability to perform fast, accurate long-haul WDM optical channel waveform modeling is vital, but the waveform modeling approach remains an open issue.

Many fiber channel models have been proposed to reduce the complexity of fiber channel modeling, such as e Gaussian noise (GN) model, nonlinear-interference noise (NLIN) model and so on. These models can quickly assess the performance of the system, such as signal-noise ratios (SNRs) and bit error ratios (BERs) [10,11,13,14]. Particularly, the universal virtual lab tool has shown speed enhancement beyond 1000 in a 45-channel dense WDM system [15]. But these works neglect the waveform accuracy in the fiber waveguide. The waveform modeling can provide full-time and full-frequency information,

This paper is submitted for review on January 17, 2022 and is supported by National Natural Science Foundation of China (62025503).

H. Yang, Z. Niu, H. Zhao, S. Xiao, W, Hu and L. Yi are with State Key Lab of Advanced Optical Communication Systems and Networks, School of Electronic Information and Electrical Engineering, Shanghai Jiao Tong University, Shanghai 200240, China (e-mail: hangyang@sjtu.edu.cn, zekunniu@sjtu.edu.cn, zhaohaochen@sjtu.edu.cn, slxiao@sjtu.edu.cn, wshu@sjtu.edu.cn, and *lilinyi@sjtu.edu.cn).

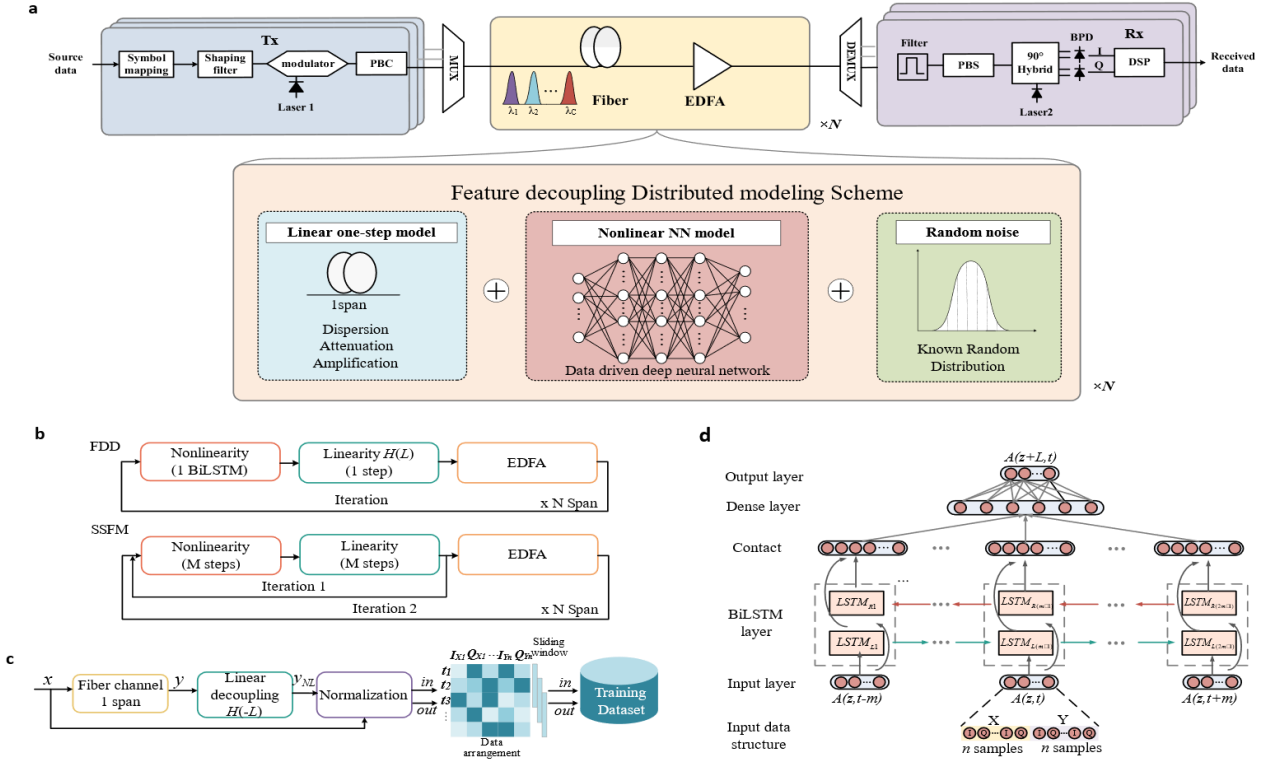

**Fig. 1. Coherent WDM System diagram and feature decoupling distributed (FDD) modeling scheme details. a,** Schematic of WDM simulation system showing the transmitter, optical fiber channel and receiver. An FDD modeling scheme replaces the fiber channel. **b,** Transmission processes of FDD-based and split-step Fourier method (SSFM)-based optical fiber channel model. **c,** Generation process for the training dataset. **d,** FDD-bi-directional long short-term memory (BiLSTM) structure for nonlinearity modeling, showing the input layer with the detailed input data structure, the BiLSTM layer, the dense layer and the output layer. The total LSTM layer number is 3. The dense layer number is 1, and no nonlinear activation function was used. The input channel data contain the current samples and the surrounding samples at other times, i.e., $A(z, t-m)$, …, $A(z, t)$, …, $A(z, t+m)$. The total time steps of BiLSTM is set at 13 in this work, that is, $m$ equals to 6.

which can be used for waveform observation, optical system optimization and digital signal processing (DSP) verification.

Deep learning techniques offer excellent nonlinear fitting capabilities and have been used to perform fast and accurate channel waveform modeling [16-18]. These works used one neural network (NN) to model the overall channel effects and can be regarded as overall modeling schemes. This purely data-driven method builds a connection between the input and output data without dependence on the theoretical model. The works cited above modelled the overall effects of single-wavelength transmission with low complexity. However, it remains challenging for overall modeling schemes to accurately capture all channel effects that occur during WDM long-haul fiber transmission. In this scenario, it becomes necessary to consider large numbers of symbols for the model inputs because of the strong inter-symbol-interference (ISI), thereby increasing the NN's complexity significantly. In addition, it is difficult for the overall modeling scheme to capture random noise levels among the spans accurately, resulting in a poor distance generalization ability. Moreover, modeling of inter-channel nonlinearities, including cross-phase modulation (XPM) and four-wave mixing (FWM) [3], is expected to be challenging. To the best of our knowledge, none of the methods above can achieve accurate waveform prediction within a short time for 40-channel long-haul fiber transmission, which is a typical optical fiber communication scenario.

Here we propose a feature decoupling distributed (FDD) modeling scheme, a hybrid model-data driven scheme, with recursion processing to model WDM long-haul fiber channels accurately and rapidly. Specifically, we initially separate the linear and nonlinear features of the channel effects and model them separately. In this case, the linearities are modelled using the model-driven methods in one step, and the NN then only learns the purely nonlinear characteristics by data-driven methods. We use bi-directional long short-term memory (BiLSTM) to model the nonlinear fiber characteristics of self-phase modulation (SPM), XPM, and FWM. Second, the proposed model focuses on one-span fiber distance fitting over a length of 80 km, and then transmits the signal recursively. The designed iterative transmission scheme can model the random noise by adding amplified spontaneous emission (ASE) noise among fiber spans, thus achieving much longer transmission distances.

The results show that the accuracy of the FDD scheme is significantly improved when compared with overall schemes. The abilities of our proposed method can be generalized to suit a wide range of scenarios, including various launch powers, multiplexed channel numbers, transmission distances and signal modulation formats. The presence of the same constellation shapes, the high overlap ratio of the waveforms, the low normalized mean square errors (MSEs), and the exact SNRs verify the high accuracy of the FDD-based BiLSTM (FDD-BiLSTM) fiber channel modeling method. The proposed method achieves 41-channel 1040-km transmission modeling within only 3 min versus more than 2 h when using SSFM. When considering the system design and optimization, it is necessary to adjust the input conditions or system parameters, and the running time of SSFM can be up to a few days, which will seriously reduce the efficiency of the channel model. Therefore, the FDD model can be an excellent and efficient channel model for multi-round system optimization. The FDD scheme highlights the remarkable achievements in WDM long-haul fiber channel modeling and opens up novel perspectives for solution of NLSE-like partial differential equations and studies of optical fiber physics.

## II. COHERENT WDM SYSTEM SETUP

To collect the training data and analyze the signal modeling performance, we simulate a coherent WDM optical transmission system, including the transmitter, the fiber channel, and the receiver. The complete system block diagram is shown in Fig. 1a and the main fiber channel parameters of this work are shown in Table 1.

We assume that the transmitter uses 16QAM as the modulation format, and that all symbols and samples are expressed as complex values in this system. Up-sampling is then implemented. The sampling rate of the system is approximately twice the number of channels multiplied by the channel baud rate, but it can also be set to other values as required. A root raised cosine (RRC) filter is used to perform signal shaping, and the roll-off factor we use is 0.1. Then, all the data are modulated into subcarriers at different frequencies for WDM as follows:

$$A(z,t) = \sum_{k=1}^{C} A_k(z,t)\exp(j\Delta\omega_k t), \quad (1)$$

where $A$ represents the optical field over two arbitrary orthogonal polarization modes $A_x$ and $A_y$, i.e., $A = [A_x, A_y]$. $A_k(z,t)$ is the envelope of channel $k$. $C$ is the total number of channels, and $\Delta\omega_k = \omega_k - \omega_0$ is the difference between the central frequency of channel k and the central frequency of the WDM comb.

In this work, polarization division multiplexing (PDM) is also considered. By considering the PMD, the polarization states can be averaged over a Poincaré sphere, and the NLSE can then derive a Manakov-PMD equation [19,20], which can be simplified as:

$$i\frac{\partial A}{\partial z} - \frac{1}{2}\beta_2\frac{\partial^2 A}{\partial t^2} + \frac{8}{9}\gamma|A|^2 A + \frac{\alpha}{2}iA = -i\frac{\Delta\beta_1}{2}\bar{\sigma}\frac{\partial A}{\partial t}, \quad (2)$$

where $A$ represents the optical field over two arbitrary orthogonal polarization modes $A_x$ and $A_y$, which represent the optical field of the polarization $x$ and $y$. $\beta_2$ is the group velocity dispersion parameter, $\alpha$ is the loss parameter, $\Delta\beta_1$ is the differential group delay caused by the PMD, and $\bar{\sigma}$ defines the mode coupling of the two local principal states of polarizations (SOP). $\gamma$ represents the nonlinear parameter. $z$ is the distance and $t$ is the time.

The SSFM algorithm divides the total transmission distance into multiple small steps for iterative simulation. Although there is no analytical expression for the overall fiber channel response, in a small transmission step, the linear and nonlinear effects can be considered to be independent of each other. The operation of each step can be expressed using:

$$A(z+h,t) \approx \exp\left(\frac{h}{2}\hat{D}\right)\exp\left\{h\hat{N}\left[A\left(z+\frac{h}{2},t\right)\right]\right\}\exp\left(\frac{h}{2}\hat{D}\right), \quad (3)$$

where $h$ is the step length, and $\hat{N}$ is the nonlinear operator, which represents the SPM, XPM, and FWM effects related to the signal energy [19]. The nonlinear effect can be expressed as:

$$A(z+h,t) = A(z,t)\exp\left(i\frac{8}{9}\gamma\left(|A_x(z,t)|^2 + |A_y(z,t)|^2\right)h\right). \quad (4)$$

The notation $\hat{D}$ represents the linear component, which denotes the effects of attenuation, chromatic dispersion (CD), and polarization mode dispersion (PMD) [21,22]. During the simulation of the linearity, the fiber of each step is represented by a concatenation of waveplates. In each of these waveplates, the operation consists of three steps. First, a rotation matrix $R(\theta,\varphi)$ is used to rotate the waveplates' SOP, and then linear effects related to the frequency are applied to the current SOP, which can be expressed as:

$$R(\theta,\psi) = \begin{pmatrix} \cos\theta & \sin\theta\exp(i\varphi) \\ -\sin\theta & \exp(i\varphi)\cos\theta \end{pmatrix}, \quad (5)$$

where $\theta$ and $\varphi$ represent the rotation angle and phase [23]. And then the effects of the CD and PMD can be implemented on the current SOP as

$$\tilde{A}_x(z+h,\omega) = \tilde{A}_x(z,\omega)\exp\left(-i\frac{\beta_2}{2}\omega^2 + i\frac{\Delta\beta_1}{2}\omega\right)h, \quad (6)$$

$$\tilde{A}_y(z+h,\omega) = \tilde{A}_y(z,\omega)\exp\left(-i\frac{\beta_2}{2}\omega^2 - i\frac{\Delta\beta_1}{2}\omega\right)h.$$

Finally, the SOP is rotated back to the original states. In principle, the rotation of SOP in each waveplate should be a random value since the random polarization state as the reviewer mentioned. But we assumed that the rotation angle and phase are both constant values of π/4 during the simulation process. This is not an accurate PMD implementation. This assumption can simplify the linear compensation algorithm and the one-step linear modeling, and the feature decoupling dataset can be easily obtained. Therefore, this approximation can apply an easily understood demo of the FDD scheme. After the channel modeling simulation by SSFM, EDFA is used to provide amplification among the fiber spans and the EDFA introduces ASE noise in the meantime [24].

The receiver uses a matched RRC filter and then uses the linear compensation algorithm to compensate CD and PMD [25]. The CD and PMD compensation process also contain three steps like the linear modeling process. First, a rotation matrix is used to rotate the SOP. Then CD and PMD are compensated together by

$$\tilde{A}_x(z-L,\omega) = \tilde{A}_x(z,\omega)\exp\left(-i\frac{\beta_2}{2}\omega^2 + i\frac{\Delta\beta_1}{2}\omega\right)(-L) \quad (7)$$

$$\tilde{A}_y(z-L,\omega) = \tilde{A}_y(z,\omega)\exp\left(-i\frac{\beta_2}{2}\omega^2 - i\frac{\Delta\beta_1}{2}\omega\right)(-L).$$

Compared with the modeling process described in Eq. (6), the compensation distance is set to $L$, i.e., the total transmission distance, and the sign is negative, representing a compensation process. Finally, the SOP is rotated back to the original states. Note that the PMD we use is an approximate algorithm, so the traditional multiple-input-multiple-output (MIMO) algorithm is not used here [26]. Following down-sampling, carrier phase recovery (CPR) [27] and demodulation are performed subsequently. Note that the CPR algorithm is not used for the compensation of the phase noise, but to rotate the signal to the symmetrical position of the origin, by which the calculated SNR value is correct and reasonable.

**Table 1. Fiber channel parameters**

| Parameters | Value |
|---|---|
| Carrier wavelength | 1550 nm |
| Symbol rate | 30G Baud/channel |
| Channel spacing | 50G Hz |
| Attenuation | 0.2 dB/km |
| Dispersion | 16.75 ps/(nm·km) |
| Differential group delay | 0.2 ps/km |
| Nonlinear refractive index | 2.6e-8 µm$^2$/W |
| Core Area | 80 m$^2$ |
| Span length | 80 km |
| Noise figure | 5 dB |
| roll-off factor of RRC | 0.1 |

### III. Principle of FDD modeling scheme

We focus on multiplexed signal propagation in the WDM fiber channel modeling approach using the FDD scheme. The training dataset can be obtained easily from the simulated WDM coherent optical transmission system, as illustrated in Fig. 1a. The FDD implementation replaces the fiber channel and erbium-doped fiber amplifier (EDFA) as a complete channel modeling tool. This FDD modeling scheme can improve the modeling performance by applying two key aspects: feature decoupling and distributed modeling.

Feature decoupling divides channel effects into linearities, nonlinearities, and random noises. As illustrated in Fig. 1a, the linear channel features can be modelled using a one-step model. This process is the same as the linear model in SSFM as described in Section II, consisting of SOP rotation, linear modeling, and SOP original rotation. The linear modeling method is given by

$$\tilde{A}_x(z+L_{span},\omega) = \tilde{A}_x(z,\omega)\exp\left(-i\frac{\beta_2}{2}\omega^2 + i\frac{\Delta\beta_1}{2}\omega\right)(L_{span}), \quad (8)$$
$$\tilde{A}_y(z+L_{span},\omega) = \tilde{A}_x(z,\omega)\exp\left(-i\frac{\beta_2}{2}\omega^2 - i\frac{\Delta\beta_1}{2}\omega\right)(L_{span}),$$

where $L_{span}$ represents one span link distance. Unlike the iterative steps used in the SSFM linear modeling process (shown in Eq. (6)), our proposed linear model of each span is calculated in only one step rather than the iterative steps. The channel nonlinear characteristics are modelled using an NN. After modeling of the linearities, the time correlations required for nonlinear modeling in the next step are shortened. Therefore, the nonlinear modeling complexity is reduced significantly. Additionally, the nonlinear features are enhanced because of the extraction of the linearities, leading to improved nonlinearity modeling accuracy. Random noises are regarded as additive noises and follow known distributions during progress through the simulation [24]. Distributed modeling represents the iterative processing operation of the linear one-step model, nonlinear NN model and random noise. Corresponding to the multiple repetitive spans used for long-haul fiber transmission, the signal is transmitted iteratively through a one-span model for arbitrary spans. The distributed modeling considers the noise inputs and avoids modeling the complete long-haul channel effects through a single NN, thus achieving distance generalization. Figure 1b provides clear comparison of the FDD modeling scheme with the traditional SSFM approach. Both modeling schemes are composed of iterative spans and the modeling of each span contains both nonlinearities and linearities, along with random noise. The SSFM requires numerous iterative steps to model the linearities and nonlinearities, but the FDD scheme requires only one step to perform accurate modeling, which allows the modeling complexity to be greatly reduced.

Pre-processing steps are required to obtain the training dataset, including feature decoupling, normalization and data arrangement, as illustrated in Fig. 1c. The NN uses only the first span channel input-output pairs as collected data. The nonlinear and linear channel data features are uncoupled via simple linear compensation to eliminate both chromatic dispersion (CD) and polarization mode dispersion (PMD). This process is the same as the linear compensation process described in Section II and can be expressed the same as Eq. (7). After compensation, the channel data only contain nonlinear characteristics, and these nonlinear data constitute the training dataset. If it is extended to the PMD effect of random birefringence, a MIMO algorithm needs to be added for PMD compensation to separate linear features; In the modeling process, it is also necessary to use a trainable butterfly filter for linear modeling. On the other hand, PMD has been verified that it has little influence on nonlinear effect, so it is also reasonable not to remove the PMD modeling [28]. FDD is also a feasible scheme, which can be adjusted according to users' needs.

The channel input $x$ and channel output $y_{NL}$ after linear decoupling must be normalized for rapid convergence and generalization. To realize generalization of input waves with different powers, we collect input-output channel pairs with multiple different powers. The channel input and output should be normalized using

$$\tilde{x} = x \cdot \sqrt{\frac{1}{\frac{1}{N_p}\sum_{i=1}^{N_p} P_i}}, \quad (9)$$

where $N_p$ represents the total number of the different power values, and $P_i$ is the power of the $i$th data group. Note that $P_i$ refers to the total power of the signal after WDM. If the power of each channel is defined as $P_{ch}$ with a unit of Watt, and the multiplexed channel number is $N_{ch}$, the total power $P_i$ can be described as $P_i = P_{ch} \times N_{ch}$.

The channel input-output complex pairs of the two polarizations should be arranged into a one-dimensional vector. Taking ISI caused by CD and PMD into consideration, we add the adjacent samples to the samples at the current time, where samples can be obtained using a sliding window. Note that the sliding window size can be reduced after linear feature decoupling because the shorter ISI length is maintained in the nonlinear dataset. The smaller window size means the less NN parameters and less computing complexity. It is found that the FDD-BiLSTM can achieve the optimal performance required when the window length is 13 in most cases. Therefore, we set the window size to 13, i.e. six past time steps, one current time step and six future time steps.

Here, the linearities are modeled using the traditional method as shown in Eq. (8) and the BiLSTM models the nonlinear characteristics. The LSTM is a type of NN that is used to process sequence data with memory [29]. The perfect internal memory ability of the LSTM motivated us to select it for channel modeling with a time memory. Because past and future samples both affect the current samples, we use the BiLSTM to capture the features from two directions. Figure 1d shows the complete BiLSTM structure. In the input layer, the signals at different times, i.e. [A(z, t-m), …, A(z, t), …, A(z, t+m)], are fed to the BiLSTM as an input sequence. A(z, t) represents the aligned one-dimensional vector at distance z and time t. m is the length of the surrounding samples, which also defines the time steps for the BiLSTM. After the input layer, the data are fed to the BiLSTM layer to realize the recurrent connection. The outputs from multiple time steps are concatenated to form a vector and this vector is then fed into a dense layer to obtain the output A(z+L, t).

The loss function used in this work is the smooth L1 loss. The epoch number is set at 300, and this number can be set to be larger for more accurate training. The weights of the NN are initialized by He initialization as described in the literature [30]. The optimizer used in this process is Adam [31]. The learning rate (LR) is 1e−3 initially and it decreases during the training process. We set the learning rate for each parameter group using a cosine annealing schedule [32]. The decreasing LR can improve the training accuracy, particularly for a nonlinear phase rotation.

## IV. RESULTS OF FDD MODELING SCHEME

### A. FDD scheme performance gain.

We varied the optical launch power per channel, the multiplexed channel number and the transmission distance to investigate the modeling performance of the proposed FDD scheme.

To demonstrate the gain in accuracy provided by the FDD modeling scheme, we compare the signal constellations of the overall models with that of our proposed FDD model. We use overall models proposed in previous works, including the BiLSTM [16], fully connected NN (FCNN) and conditional generative adversarial network (CGAN) models [17], which are simplified as the overall-BiLSTM, overall-FCNN and overall-CGAN, respectively. We tune the NN parameters to ensure that they are suitable for WDM channel modeling. The channel conditions are set at 3 channels, 5 dBm/channel and 800 km, and the overall models are used to model the 800 km link using only one NN. The window size required for the overall models is 501, while that required for the FDD is only 13.

The results in Fig. 2a are from central channel output after linear compensation, including CD and PMD compensation. The constellation of FDD-BiLSTM is similar with that of SSFM, which indicates the accurate modeling of channel nonlinearity and noise. The inaccurate constellation shapes show that the overall models fail to model the noise and the nonlinear effects accurately. Typically, the overall-BiLSTM and overall-FCNN are deterministic fitting models and cannot model random noises. The generative model, i.e. the overall-CGAN, can fit random distributions but it is difficult to achieve convergence. Therefore, the overall-GAN cannot generate accurate levels for the noise and nonlinearities of long-haul WDM transmission. Moreover, the constellations of FDD-FCNN and overall-FCNN neglect the phase rotation related to signal power comparing with that of BiLSTM. The results illustrate that the BiLSTM structure offers advantages over the FCNN because of the BiLSTM's internal cyclic memory structure, which is more suitable for memory channel applications.

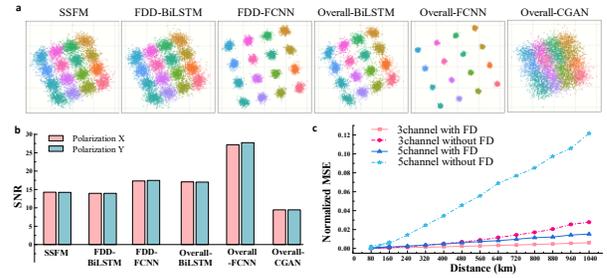

Fig. 2. Performance gain of FDD scheme compared to other methods. **a,** Central channel output constellations after the compensation of chromatic dispersion (CD) and polarization mode dispersion (PMD) with different models for 3 channels, 5 dBm/channel, and 800 km. The results are all from the X polarization. **b,** SNRs of central-channel output generated using different models corresponding to that in **a**. **c,** Normalized MSEs of distributed BiLSTM with or without feature decoupling (FD) for 3channels and 5chanels with 5dBm/channel.

We also calculate the signal-to-noise ratio (SNR) of the signals generated by above different models and then processed by the same DSP to allow a quantitative comparison to be made. The SNR is defined as

$$SNR = 10\log_{10}\left(\frac{P_s}{\mathbb{E}|rx-tx|^2}\right), \quad (10)$$

where $rx$ and $tx$ are the received and transmitted samples, respectively, $P_s$ is the signal power, and represents the calculation of mean values. We measure the SNR of the central channel output after dispersion compensation and CPR; the SNR thus represents the nonlinear noise level and the ASE noise level under the same signal power conditions. The SNR difference of FDD-BiLSTM and SSFM is within 0.5dB, but the SNR difference of other models ranges from 3dB to 13dB. The incorrect SNRs from the overall models shown in Fig. 2b verify the failure of the overall methods for WDM fiber channel modeling. The similar constellation and coincident SNR of the FDD-BiLSTM model demonstrate its strong ability for fiber channel modeling when compared with the overall methods. In addition, the similar signal performances after the same DSP verify that the FDD-BiLSTM channel model can be used for fast algorithm verification.

The distributed BiLSTM's performance with and without feature decoupling (FD) is compared. To enable direct quantitative comparison of the modeling errors, the normalized MSE [16] is calculated as follows:

$$Nor\_MSE = \frac{\sum_{i=1}^{N_{data}}|\hat{y}_i - y_i|^2}{\sum_{i=1}^{N_{data}}|y_i|^2}, \quad (11)$$

where $N_{data}$ is the number of data size, $y$ and $\hat{y}$ are the SSFM output and the data generated by the distributed NN model, respectively. Note that ASE noise should not be added among the spans when recording the normalized MSE, because the random noise will lead to incorrect values.

Figure 2c illustrates the normalized MSE of distributed BiLSTM with or without FD for 3channels and 5chanels with 5dBm/channel vs. distance characteristics. The normalized MSE of BiLSTM without FD increases quickly with increasing distance, especially for large multiplexed channels, which indicates that a severe cumulative error is caused by the iteration. In contrast, the BiLSTM with FD reduces the cumulative error effectively because the NN only needs to learn the nonlinear features. The linear features are modelled in the traditional model-driven manner, which ensures the accuracy of the linear features. In addition, the nonlinear features are strengthened after decoupling, which results in easy and accurate fitting.

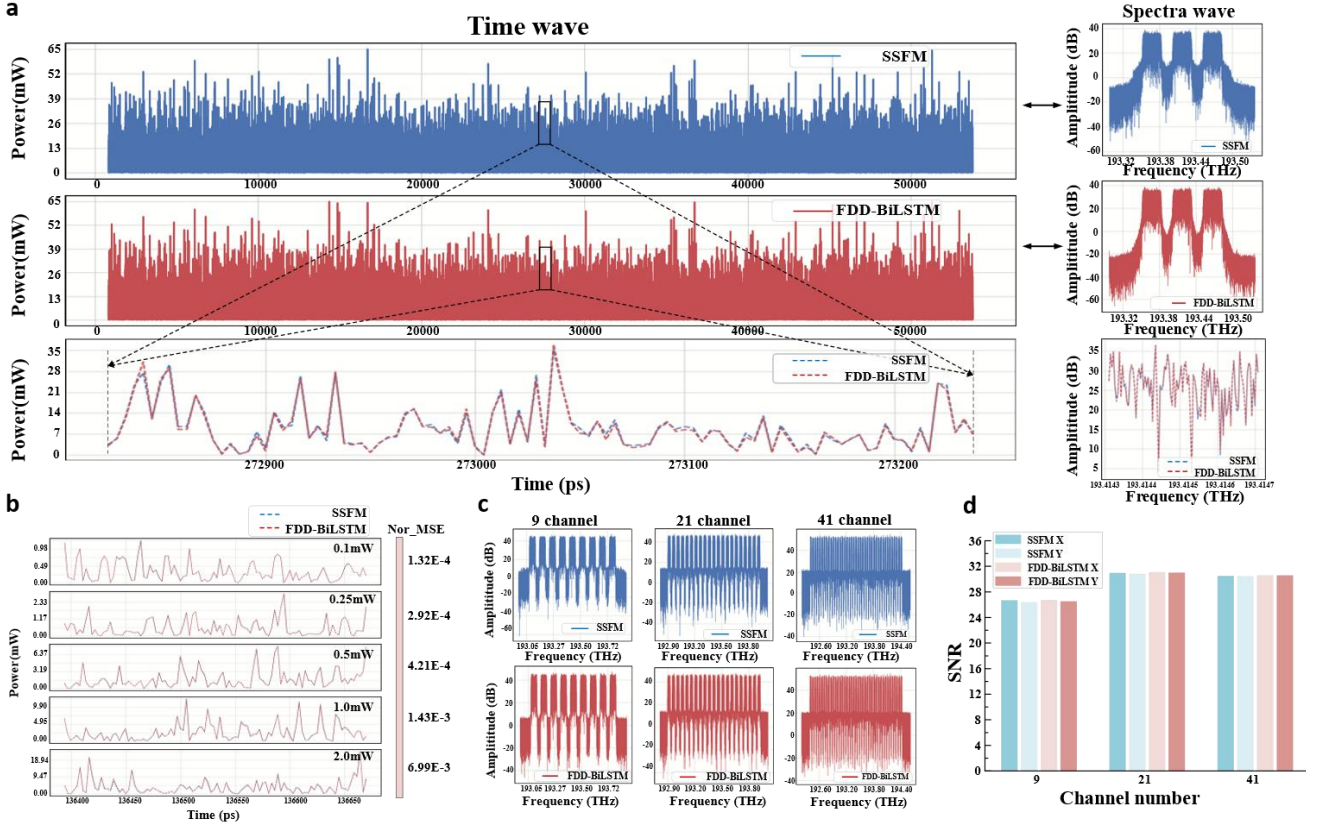

**Fig. 3 FDD scheme accuracy under different conditions. a,** Time wave and spectra wave characteristics of the channel output generated by the SSFM and FDD scheme for 3 channels, 5 dBm/channel, and 800 km. **b,** Time waves of SSFM output and FDD-BiLSTM output for 5 channels, 400 km, and 0.1, 0.25, 0.5, 1.0, 2.0mW/channel respectively, which are contained in the training dataset. **c,** Optical spectra of channel outputs based on SSFM and FDD-BiLSTM for 9, 21, 41 channels with 5dBm, -1dBm and -1dBm per channel respectively at 80km. **d,** SNRs of channel output based on SSFM and FDD-BiLSTM under the same conditions as **c**.

### B. The modeling Accuracy

We demonstrate the fiber channel modeling accuracy of the FDD-BiLSTM from the perspectives of long transmission distances, different launch powers, and different WDM channel numbers.

The optical time and spectra waveforms simulated using SSFM and FDD-BiLSTM under transmission conditions of 3 channels, 5 dBm/channel and 800 km distance are presented in Fig. 3a. In general, the channel outputs generated by the SSFM and FDD-BiLSTM are extremely similar in both the time and frequency domains. Specifically, we magnify the middle sections of these waveforms, which show a great deal of overlap. The high degree of time-frequency consistency demonstrates the modeling accuracy of the FDD-BiLSTM for long transmission distances.

The performances for different optical launch powers are also investigated under transmission conditions of 5 channels and 400 km distance. Figure 3b illustrates the optical time waveforms of the channel outputs based on the SSFM and FDD-BiLSTM for various powers. The normalized MSEs at 0.1, 0.25, 0.5, 1.0 and 2.0 mW/channel are 1.32E−4, 2.92E−4, 4.21E−4, 1.43E−3 and 6.99E−3, respectively. The normalized MSEs increase with increasing launch powers because growing distortions are caused by the high-intensity nonlinearity. However, these normalized MSE values are still at a low level. The remarkable similarities and low normalized MSEs illustrate the high accuracy of FDD-BiLSTM modeling for various launch powers.

The results of modeling of different numbers of WDM channels are shown in Fig. 3c for 9, 21 and 41 channels. The launch powers are set 5dBm, -1dBm, -1dB/channel respectively with 80km distance. The results show high spectral consistencies for the different channel numbers between the SSFM and FDD methods. The similar SNRs of both polarizations in Fig. 3d also show the modeling accuracy for the different numbers of channels.

### C. Generalization.

The generalization process is essential for practical channel modeling applications, particularly for different transmission conditions and a variety of inputs, including powers, the numbers of WDM channels and the modulation formats.

We use FDD-BiLSTM trained under the conditions in Fig. 3b to test power generalization ability. During the testing process, the power is set to 0.175, 0.375, 0.75 and 1.5 mW/channel respectively that never occur in the training dataset. The time waveforms are shown in Fig. 4a, and the normalized MSEs are 2.28E−4, 3.80E−4, 1.53E−3 and 3.21E−3, respectively. The same trends for these MSEs are found in the testing and training datasets. The low normalized MSEs illustrate the generalization of the FDD-BiLSTM under different optical launch power conditions.

We then realized generalization of the different numbers of WDM channels. Because of the various sampling numbers used for the different numbers of channels, the input layer dimension is not fixed, and this is difficult for a model to adapt to all channels. Nevertheless, we can use a model suitable for a large number of channels to work for a small number of channels by transmitting all-zero signals in some channels. The results in Fig. 4b show the optical spectra of the central channel for different numbers of WDM channels. The FDD-BiLSTM is trained under the 5-channel condition. The signal of first channel, second channel, second and fourth channels are set to 0 respectively during the testing process. The similar spectra waves illustrate the channel number generalization produced by setting the all-zero signals at some channels.

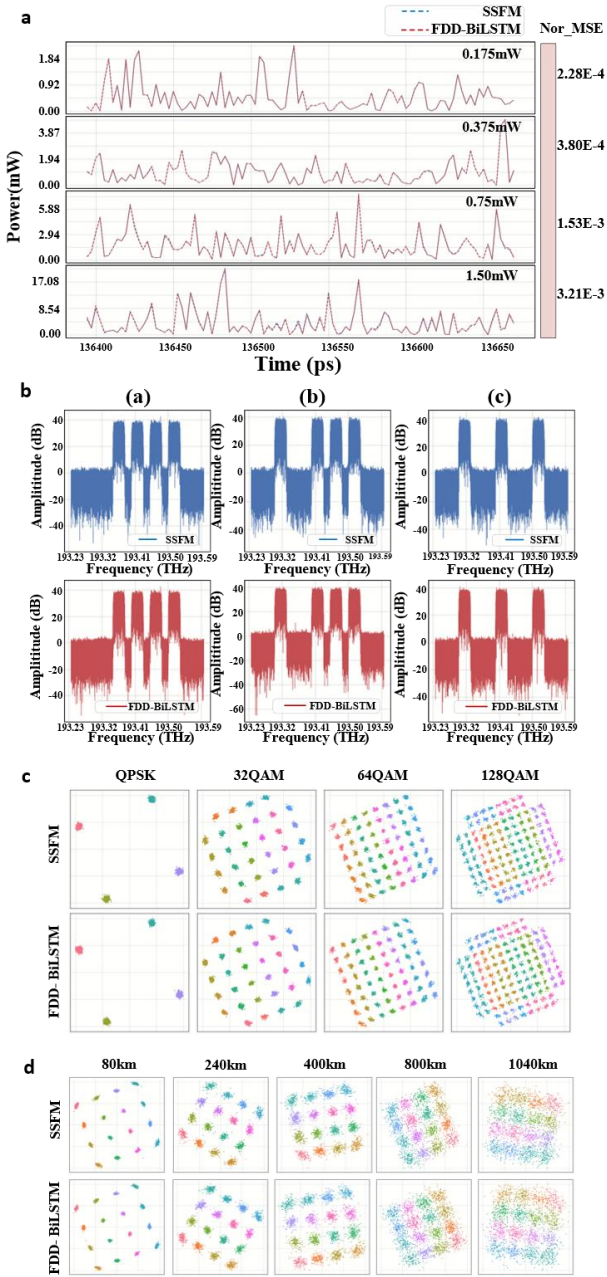

**Fig. 4. Generalization of FDD scheme for different conditions. a,** Time waves of SSFM and FDD-BiLSTM outputs for 5 channels, 400km, 0.175, 0.375, 0.75 and 1.50 mW/channel respectively, which are not contained in the training dataset. **b,** Optical spectra of channel outputs based on SSFM and FDD-BiLSTM for 3dBm/channel and 80km. The model is trained at 5 channels, but the transmitted signals of (a) the first channel, (b) the second channel, (c) the second and fourth channels are set to 0 in the testing process. **c,** Constellations of SSFM and FDD-BiLSTM outputs after linear compensation with different modulations for 5 channels, 3 dBm/channel and 80 km. **d,** Constellations of SSFM and FDD-BiLSTM outputs after linear compensation with different transmission distances for 3 channels with 5 dBm/channel.

During training, the training dataset contains only 16 quadrature amplitude modulation (QAM) symbols. We varied the test dataset inputs to use quad-phase shift keyed (QPSK) modulation along with 32QAM, 64QAM and 128QAM. As shown in Fig. 4c, the FDD-BiLSTM outputs after linear compensation show a good generalization ability for the different modulation formats, regardless of whether they are high- or low-order. We also calculated normalized MSEs for the channel output without random noise. The MSEs of the inputs with QPSK, 32QAM, 64QAM and 128QAM are $1.04E{-}4$, $1.47E{-}4$, $1.66E{-}4$ and $1.50E{-}4$, respectively, which are similar to the results obtained for 16QAM under the same conditions. The output constellations also demonstrate the generalization ability for the transmission distances, as shown in Fig. 4d. The training dataset only contains input-output pairs for the 80 km link, and the trained FDD model can be transmitted iteratively for the long-haul fiber link.

Through selection and pre-processing of the datasets and appropriate modeling scheme design, the model has strong generalization abilities for the launch powers, the numbers of WDM channels, the modulation formats and the transmission distances. There are different reasons for generalization of these conditions.

• Discrete feature independence. The NN model input is the sampled data after up-sampling. Therefore, the input is equivalent to the analogue waveform input, which makes the NN model independent of the discrete data characteristics. The modulation format and the other discrete characteristics can be generalized easily by the NN.

• Signal parameters visualization. The essence of generalization of the launch powers and channel numbers is that the NN can capture the input amplitude differences and frequency components well. The data normalization method ensures the amplitude differences for signals with different input powers. The multiplexed signal is used to construct the training dataset and the frequency information corresponding to the different channel numbers is contained directly in the input signal. Therefore, the NN model has both power and channel number generalizations.

• Distributed modeling scheme. Because the proposed distributed modeling scheme aims to divide the long-distance transmission channel into multiple spans, we can then realize generalization of different distances flexibly.

### D. Complexity

Next, we compare the complexities of the SSFM and FDD-BiLSTM in terms of their running times. The codes for the different model schemes run on the same server using NVIDIA GeForce RTX 3090 Computer Graphics Cards. Here we consider two commonly used SSFM algorithms: the constant step size algorithm and the variable step size algorithm. The constant step size in the former is set to 0.01 km per step, while the variable step size algorithm is based on a nonlinear phase rotation limitation, and the maximum nonlinear phase shift value is 0.005 rad for each step [33] Therefore, we call these algorithms the constant-step-size SSFM (C-SSFM) and the nonlinear phase SSFM (NP-SSFM) for convenience. The C-SSFM is an accurate fibre channel modeling method and the NP-SSFM is an approximate method used to accelerate the running speed. We record the running time characteristics versus both transmission distance and channel numbers for different algorithms. Because of the different up-sampling numbers used, the data size is united using symbol numbers. The following results are all based on the 16384 symbols transmitted for each channel and polarization.

The execution times for different transmission distances on the graphics processing unit (GPU) and central processing unit (CPU) were recorded as shown in Fig. 5a. We selected the model for 15 channels and 5 dBm/channel. The running time for the C-SSFM is the longest, that of the NP-SSFM is shorter, and that for the FDD-BiLSTM is the shortest. The long-haul fiber transmission time increased linearly with increasing distance for both SSFM and NN because of the iterative transmissions in both cases. The C-SSFM simulation method takes approximately 398 s on the GPU and 2588 s on the CPU at the 1040 km distance. The NP-SSFM method takes 98 s on the GPU and 662 s on the CPU. As a significant contrast, the FDD-BiLSTM takes only 12 s on the GPU and 45 s on the CPU. When compared with the accurate C-SSFM method, the running speed of the FDD method on the CPU is nearly 60 times higher.

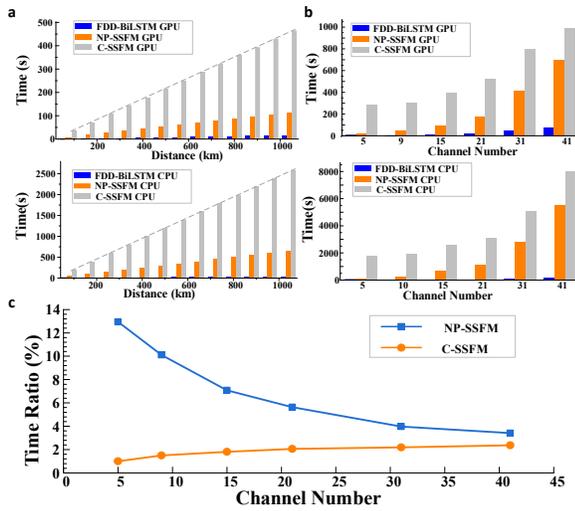

**Fig. 5. Complexity comparison of FDD scheme and SSFM. a,** Running times for NP-SSFM, C-SSFM and FDD-BiLSTM under 15 channels, 5 dBm/channel and 80–1040 km conditions on the GPU and CPU. **b,** Running times of NP-SSFM, C-SSFM and FDD-BiLSTM for 5, 9, 15, 21, 31 and 41 channels, 5 dBm/channel and 1040 km on the GPU and CPU. **c,** Time ratios of FDD-BiLSTM to NP-SSFM and C-SSFM at different channel numbers on the CPU.

We also recorded the running times and analysed the performance for different channel numbers. The transmission power in each channel is set at 5 dBm and the transmission distance is 1040 km. When the number of channels is increased, the up-sampling number on the signal also increases. Therefore, the simulation running time for the channel transmitting the same symbols also increases, as shown in Fig. 5b. The running times of the C-SSFM, NP-SSFM and FDD-BiLSTM algorithms are 288.3 s, 22.5 s and 2.6 s for 5 channels on the GPU, respectively. For 41 channels and under the CPU conditions, the C-SSFM and NP-SSFM algorithms take 8020 s and 5564 s, i.e. 2.23 h and 1.55 h respectively. The extremely long time required for the SSFM seriously reduces its research efficiency and hinders fiber channel model application. The time required for the FDD-BiLSTM under the same conditions is only 190 s, i.e. 3.17 min, which is far shorter than that for the SSFM.

To analyse the complexity optimization performance, we calculate the time ratio to represent the degree by which the time is reduced; this ratio is defined as

$$time\_ratio = \frac{time_{NN}}{time_{SSFM}} \times 100\%. \quad (12)$$

The ratios of the FDD-BiLSTM to the NP-SSFM and C-SSFM algorithms for the different channel numbers on the CPU are shown in Fig. 5c. The minimum time ratio reaches 1% for 5 channels. With increasing numbers of channels, the complexity optimization space becomes larger when compared with that of the NP-SSFM and smaller when compared with that of the C-SSFM.

• When the number of channels increases, more processed samples are then required. Therefore, the complexity of the C-SSFM will also increase. However, regardless of how high the input power becomes, the total number of iteration steps is fixed, which means that the C-SSFM complexity increases slowly.

• With increasing numbers of channels, the multiplexed signal power also increases. The NP-SSFM step size then needs to be smaller. Therefore, the total number of iterative steps will also increase, leading to a rapid increase in complexity.

• When the number of channels increases, more processed samples are then required. The data dimensions thus increase for the input and hidden layers of the FDD-BiLSTM model, resulting in increased complexity.

The complexity of the FDD-BiLSTM increases more rapidly than that of the C-SSFM but more slowly than that of the NP-SSFM, meaning that the ratios of the two SSFM models change with the number of channels. For 41 channels, the running time ratio of the proposed method gives approximately 3% complexity for both the C-SSFM and NP-SSFM. The results show that the NN running time can be reduced significantly when compared with that of the SSFM. In addition, even for smaller step lengths, the SSFM complexity will increase to a greater extent, but the complexity will remain constant for the NN-based model.

## V. CONCLUSION

We have demonstrated how the FDD scheme can accurately and rapidly achieve the simulation required for WDM long-haul fiber channel modeling. This simulation is accomplished by performing recursion processing and combining the NLSE-derived model-driven method for one-step construction of linearities with the data-driven approach for nonlinear fitting, where the BiLSTM is competent in providing the nonlinear representation.

When compared with the overall methods, the recursion model brings notable advantages for modeling of noise and nonlinearities in terms of long-haul fiber transmission. Essentially, the noise-signal interactions are captured naturally by the iterative transmissions with additive ASE noise. This advantage increases for the distance generalization procedure that requires the NN to simulate noise variations accurately. Additionally, the feature decoupling concept presented here uses expert knowledge capabilities to develop an efficient framework for the nonlinearity-focusing NN. In this case, the NN task for high ISI modeling is eased, and there is no need to use a multi-split-step method to build a connection between the linearities and nonlinearities. As a result, the complexity of the framework presented here is reduced substantially. We have demonstrated the high accuracy, strong generalization ability, and ultrafast operating speed of the FDD modeling scheme, achieving 41-channel 1040-km fiber transmission modeling with only 3% of the time complexity of the SSFM. In further research, this scheme can also be extended to other scenes, such as the subsea long-haul link, but the modeling errors at the long distances should be reduced by further explorations.

The proposed FDD modeling scheme provides accurate and rapid prediction of signal evolution in WDM fiber channels. This simulation tool can accelerate advancement of research into fiber nonlinearities required to explore the channel capacity of nonlinear fibers. This computational approach is also conducive to fiber communication system optimization, breaking through the physical limitations of expensive WDM system devices. Furthermore, this work acts as a reference for numerical solution of NLSE-like nonlinear partial differential equations from a combined data-driven and model-driven perspective.